\documentclass[showpacs, prb,twocolumn,preprintnumbers , superscriptaddress, aps]{revtex4-2}

\usepackage{color}
\usepackage{amsmath,amssymb}
\usepackage{pifont}
\usepackage{amssymb}  % More symbols
\usepackage{bbold}
\usepackage{float}

\usepackage[labelfont=bf]{caption} 
\usepackage{tikz}
\usepackage{makecell}
\usepackage{subfig}
\usepackage{pifont}   % Ding symbols
\usepackage{graphicx} % Include figure files
\usepackage{dcolumn}  % Align table columns on decimal point
\usepackage{bm}       % bold math
\usepackage{multirow} % Table functions
\usepackage{placeins}% For making floats not move around everywhere
\usepackage[colorlinks]{hyperref}
\usepackage{mathtools}

\newcommand{\vect}[1]{\mathbf{#1}}
\newcommand{\ket}[1]{\left|{#1}\right\rangle}
\newcommand{\bra}[1]{\left\langle{#1}\right|}

\newcommand{\abs}[1]{\left\|{#1}\right\|}

\usepackage{breqn}
\usepackage{mathrsfs}

\usepackage{multirow}

\begin{document}

\title{Topological transverse spin transport in a canted antiferromagnet/heavy metal heterostructure}
    \author{Wesley Roberts}
%	\email{roberts.w@northeastern.edu}
	   \affiliation{
        Department of Physics$,$ Northeastern University$,$ Boston MA 02115$,$ USA
        }
    \author{Bowen Ma}
    \affiliation{%
    Department of Physics and HK Institute of Quantum Science \& Technology$,$ The University of Hong Kong$,$ Hong Kong$,$ China
    }
    \author{Martin Rodriguez-Vega}
    \affiliation{%
    Department of Physics$,$ The University of Texas at Austin$,$ Austin$,$ TX 78712$,$ USA
    }
    \affiliation{
        Department of Physics$,$ Northeastern University$,$ Boston MA 02115$,$ USA
        }
    \author{Gregory A. Fiete}
     \affiliation{
        Department of Physics$,$ Northeastern University$,$ Boston MA 02115$,$ USA
        }
    \affiliation{Department of Physics$,$ Massachusetts Institute of Technology$,$ Cambridge$,$ MA 02139$,$ USA}

\begin{abstract}
    We theoretically study the conditions under which a spin Nernst effect - a transverse spin current induced by an applied temperature gradient - can occur in a canted antiferromagnetic insulator, such as  ${\rm LaFeO_3}$ and other materials of the same family. The spin Nernst effect may provide a microscopic mechanism for an experimentally observed anomalous thermovoltage in ${\rm LaFeO_3}$/Pt heterostructures, where spin is transferred across the insulator/metal interface when a temperature gradient is applied to ${\rm LaFeO_3}$ parallel to the interface [W. Lin ${\it et \; al}$, Nat. Phys. ${\bf 18}$, 800  (2022)]. We find that ${\rm LaFeO_3}$ exhibits a topological spin Nernst effect when inversion symmetry is broken on the axes parallel to both the applied temperature gradient and the direction of spin transport, which can result in a spin injection across the insulator/metal interface. Our work provides a general derivation of a symmetry-breaking-induced spin Nernst effect, which may open a path to engineering a finite spin Nernst effect in systems where it would otherwise not arise.
\end{abstract}

\maketitle

\section{Introduction}
The study of spin transport phenomena is vital to the growing field of spintronics \cite{RevModPhys.76.323, spintronicsreview}, which seeks to use the spin degree of freedom to transfer energy or information, analogous to the use of electron charge in electronics. In particular, spin caloritronics considers thermally driven spin and heat flow in magnetic systems \cite{Bauer2012-dl, YU2017825,Lujan2022-yd}. One area of growing interest in this field is magnonics \cite{YUAN20221,Barman_2021, Kruglyak_2010}, in which the relevant spin transport is mediated by magnons \cite{doi:10.1146/annurev-conmatphys-031620-104715,Dos_Santos_Dias2023-gs}--low-energy magnetic fluctuations that manifest as excitations above a magnetically ordered ground state. Magnons can exhibit topological transport effects such as a topological thermal Hall effect \cite{magnonthermalhall, PhysRevB.95.165106,PhysRevLett.107.236601,Onose2010-bw,PhysRevLett.104.066403,PhysRevLett.106.197202,PhysRevB.85.134411,PhysRevB.87.144101,PhysRevB.89.134409,PhysRevB.98.094419}, where spin currents are driven by thermal gradients and deflected by Berry curvature effects \cite{PhysRevB.105.L161103}. However, due to the bosonic statistics of magnons, the thermal responses of magnons are not quantized (due to the absence of ``filled bands") as they are in the electronic analog where the Fermi statistics allow for a notion of a ``filled band." The study of magnon systems also has important technological potential originating in the transport of magnons as well as cavity magnonics \cite{ZARERAMESHTI20221}.

In this work, we theoretically explore spin transport in the antiferromagnetic insulator ${\rm LaFeO_3}$ (LFO) \cite{Park_2018, C7RA06542F,10.1093/micmic/ozad067.842}, whose crystal structure and magnetic order are shown in Fig.\ref{Fig:LFOdiagram}. LFO has a perovskite structure \cite{Bousquet_2016} with space group Pbnm. Below its transition temperature, it has noncollinear antiferromagnetic order, with small spin canting giving rise to weak ferromagnetism along the $c$-axis \cite{PhysRevB.102.104420}. This weak ferromagnetism arises due to spin-oribit coupling effects that underlie anisotropy terms in the magnetic Hamiltonian \cite{DZYALOSHINSKY1958241, Moriya}. In our work, we explore a microscopic mechanism that could explain an experimentally observed anomalous thermovoltage in LFO/Pt heterostructures \cite{LFOExp}, where spin is transferred across the insulator/metal interface when a temperature gradient is applied to LFO {\em parallel} (and also perpendicular) to the interface. This spin transport is consistent with a spin Nernst effect (SNE) \cite{PhysRevResearch.2.013079,PhysRevLett.117.217203, PhysRevB.93.161106, PhysRevLett.117.217202,Zhang2022-cw,PhysRevResearch.4.013186,He2021-yv,PhysRevB.104.174410} in bulk LFO - however, under normal conditions, LFO does not support a SNE. We show that under sufficient lowering of the symmetries of LFO, namely a loss of inversion symmetry along the directions of both the applied temperature gradient and the normal to the interface, a nonzero SNE arises. We argue that one way in which this lowered symmetry could manifest is through a weak magnetic dimerization along each of these axes.  Although our work is motivated by ${\rm LaFeO_3}$, our analysis is more general and applicable to a broader class of antiferromagnetic materials with small canting. 

We begin our discussion in Sec.\ref{Sec:experiment} by briefly summarizing the observed transport phenomenon for which we propose a model.  In Sec.\ref{Sec:bulk}, we elaborate on the minimal model for LFO: We present its low-energy excitations in the form of magnons using a Holstein-Primakoff expansion. The magnon dispersion, Berry curvature, and spin Berry curvature are presented. We focus on aspects of the model that enforce a zero SNE. Next, in Sec.\ref{Sec:symmetry}, we explore what constraints must be relaxed in order to achieve a nonzero SNE. The relevant symmetries prohibiting a finite SNE manifest most obviously in the spin Berry curvature. We present a Hamiltonian term that breaks these symmetries, and in Appendix \ref{App:HD}, we show that this is the most general form of such a term. We then analyze the new magnon band structure and spin Berry curvature for the symmetry-broken model and present the temperature dependence of the (nonzero) SNE. Finally, in Sec.\ref{Sec:conclusion}, we conclude our discussion and provide an outlook for further work.

\section{Experimental transport phenomenon} \label{Sec:experiment}
We begin with a brief discussion of the experimental setup that motivates our theoretical study \cite{LFOExp}. In the experiment, a conductor with large spin-orbit coupling (Pt or W) was placed on a sample of LFO, an antiferromagnetic insulator, to which was applied a temperature gradient parallel to the interface (as well as perpendicular, but the more standard perpendicular orientation is not our focus here). Various directions and strengths of magnetic field were also applied, but the primary result, regardless of the magnetic field configuration, was a magneto-thermovoltage observed in the conductor, ultimately due to spin pumping across the interface, which was detected as a voltage via the inverse spin Hall effect (hence the need for large atomic number metallic elements) \cite{ISHE, PhysRevLett.97.216603, PhysRevLett.98.156601}. 

\begin{figure}[htp]
    \centering
\includegraphics[width=8cm]{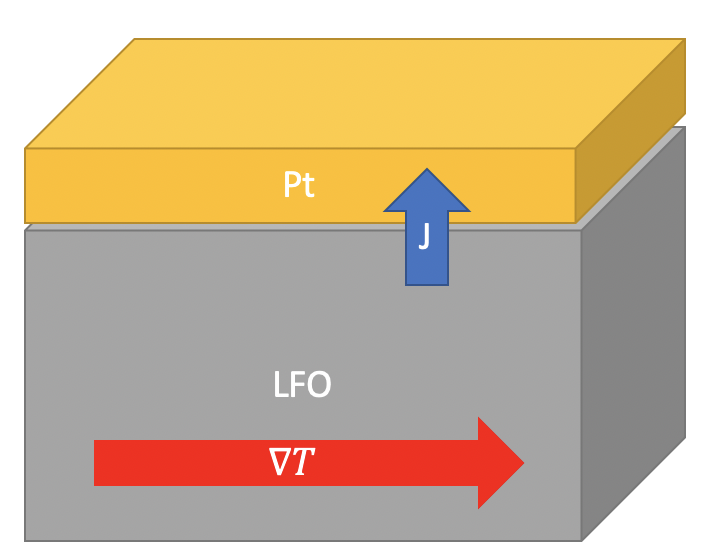}    \caption{\label{Fig:experiment}Schematic of the experimental setup of Lin et al. \cite{LFOExp}, for which we discuss a microscopic mechanism. Pt is a heavy metal ``capping layer" for LaFeO$_3$ (LFO).  The red arrow indicates the direction of the temperature gradient $\nabla T$ and the blue arrow indicates the direction of the spin current $J$.}
    \end{figure}

The geometry of the device is depicted in Fig.\ref{Fig:experiment}. Note that a transverse spin current across the interface is not experimentally observed when LFO is replaced with a ferromagnetic insulator, unless a component of the temperature gradient is directed perpendicular to the interface to induce a longitudinal spin Seebeck effect\cite{Uchida_2014}.

We propose the spin Nernst effect (SNE) \cite{PhysRevB.93.161106} as a candidate microscopic explanation for spin transport across the interface in the parallel temperature gradient geometry shown in Fig.\ref{Fig:experiment}. In the SNE, a transverse spin current with a given polarization is induced in the presence of a temperature gradient, so that $J^{\mu}_{\nu} = \alpha^{\mu}_{\nu \beta} \nabla_{\beta}T$ where all the greek indicies take the values $x,y$ or $z$. The SNE amounts to cases in which $\alpha^{\mu}_{\nu \beta} \neq 0$ for $\nu \neq \beta$. However, as we discuss in the next section, bulk LFO does not on its own produce a nonzero SNE. Instead, it does so when its symmetry is sufficiently lowered - namely when inversion symmetry is broken along the $\nu$ and $\beta$ directions. This is detailed in Sec.\ref{Sec:symmetry}.

\section{Properties of bulk ${\rm LaFeO_3}$} \label{Sec:bulk}

\begin{figure}[htp]
    \centering
\includegraphics[width=8cm]{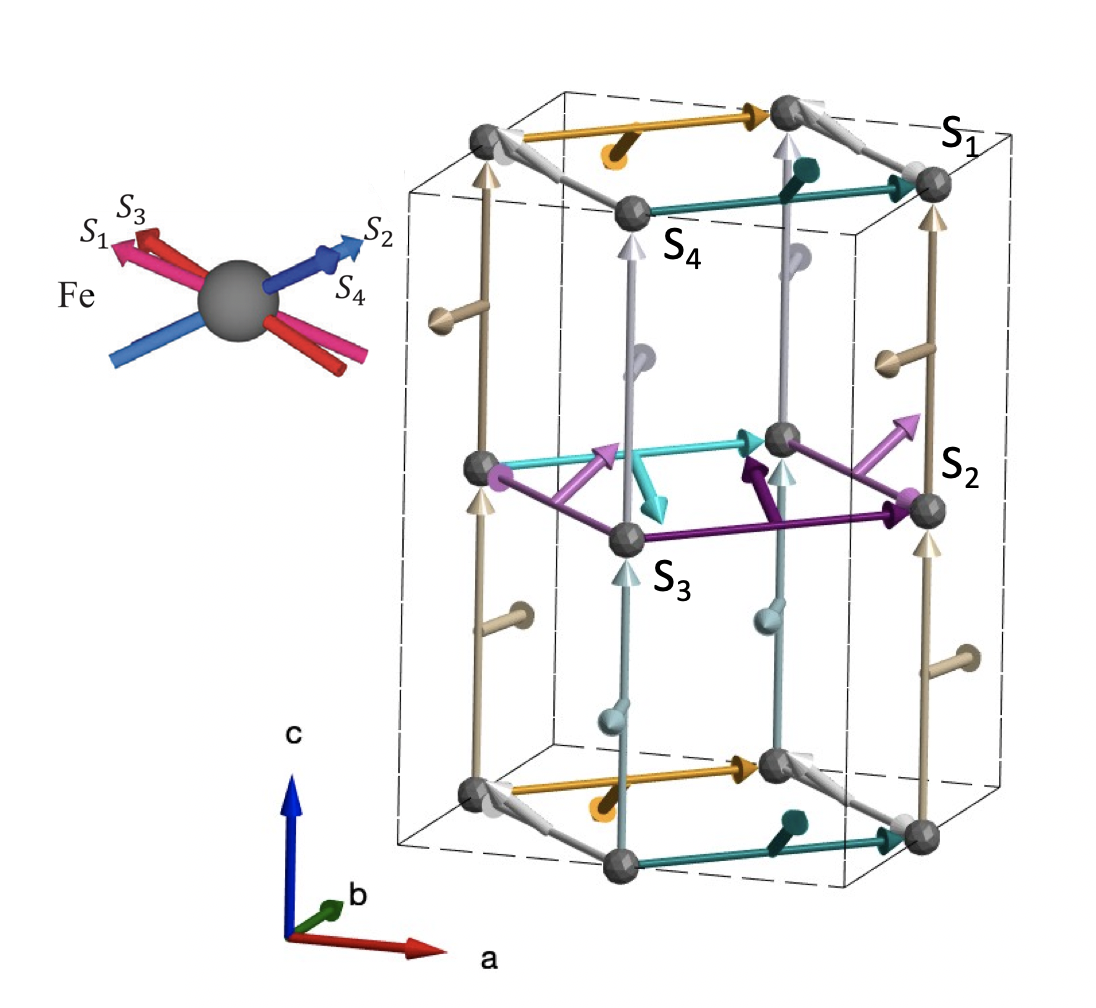}   \caption{\label{Fig:LFOdiagram}Ground state configuration of LFO without an applied field, with DM coupling vectors depicted as arrows along the bonds between the spins. The $c$-axis is ferromagnetic (has a small net moment), while the $a$- and $b$-axes are antiferromagnetic (have nearly cancelling moments).}
    \end{figure}

${\rm LaFeO_3}$ (Fig.\ref{Fig:LFOdiagram}) is a member of the rare-earth orthoferrites \cite{Park_2018,C7RA06542F,COUTINHO201759, 10.1063/1.1657530, PhysRev.125.1843, KOEHLER1957100}. The Fe atoms have spin $5/2$, inviting an analysis of spin fluctions in the language of magnon quasiparticles. The transition temperature of LFO is 738K \cite{Park_2018}, and is therefore a magnetically ordered insulator at room temperature. We take as a starting point a Hamiltonian with nearest-neighbor Dzyaloshinsky-Moriya interactions (DMI) \cite{DZYALOSHINSKY1958241,Moriya} mediated by the oxygen atoms lying between neighboring Fe atoms. Contrary to the structure of other perovskites, the Pbnm structure of LFO supports a non-vanishing DMI, where only canted G-type AFM order is allowed \cite{LFOExp}. The existence of a DMI in LFO is responsible for the small canting in the magnetic order which ultimately allows for nontrivial Berry curvature and spin Berry curvature in the magnon bands, discussed in detail below.

The minimal magnetic Hamiltonian for LFO is given by \cite{Park_2018}:
\begin{dmath} \label{eq:LFO}
    H = J_c\sum_{{\rm along}\: c} \vec{S}^L_i \cdot \vec{S}^L_j + J_{ab}\sum_{ab \: {\rm plane}} \vec{S}^L_i \cdot \vec{S}^L_j + J'\sum_{\langle \langle i,j \rangle \rangle} \vec{S}^L_i \cdot \vec{S}^L_j + \sum_{\langle i,j \rangle} \vec{D}_{ij} \cdot \vec{S}^L_i \times \vec{S}^L_j + K_a \sum_i (S^{Lx}_i)^2 + K_c \sum_i (S^{Lz}_i)^2 + \sum_i \vec{h}\cdot \vec{S}^L_i,
\end{dmath}
where the $L$ superscript indicates that the spins are written in a global ``lab" frame, with the $a$, $b$, and $c$-axes corresponding to the $x$, $y$, and $z$ directions, respectively. The Fe atoms have a $S = 5/2$ moment. Here $J_c$ is the nearest-neighbor magnetic exchange energy along the $c$-axis, $J_{ab}$ the nearest-neighbor magnetic exchange in the $ab$-plane, $J'$ the second-nearest neighbor exchange, $\vec D_{ij}$ the Dyzaloshinskii-Moriya interaction between site $i$ and $j$, $K_a$ the Ising anistropy in the $ab$-plane along the $x$-direction, $K_c$ the Ising anistropy along the $c$-axis in the $z$-direction and $\vec h$ is an externally applied magnetic field present in the experiments \cite{LFOExp}. The parameters in Eq.\eqref{eq:LFO} have been found via inelastic neutron scattering \cite{Park_2018}, and are collected in Table \ref{Tab:params}. 

$\newline$
\begin{table} 
\begin{tabular}{ |c|c| }
 \hline
 \multicolumn{2}{|c|}{LFO parameters (meV)} \\
 \hline
 $J_c$   & 5.47   \\ \hline
 $J_{ab }$ & 5.47\\ \hline
 $J'$ & 0.24 \\ \hline
 $D_{ab}$    & 0.130 \\ \hline
 $D_c$ &  0.158 \\ \hline
 $K_a$ & -0.0124 \\ \hline
 $K_c$ & -0.0037 \\ 
 \hline

\end{tabular}

\qquad

\begin{tabular}{|c|c|c|c|}
\hline
 \multicolumn{4}{|c|}{$\vec{D}_{ij} = D_{ij}(\alpha_{ij},\beta_{ij},\gamma_{ij})$} \\
   \hline 
    & $\abs{\alpha}$ & $\abs{\beta}$ & $\abs{\gamma}$\\ \hline $D_{ab}$  & 0.554 & 0.553 & 0.623  \\ \hline
    $D_{c}$ & 0.191 & 0.982 & 0 \\ \hline 
\end{tabular}

\caption{List of LFO parameters appearing in Eq.\eqref{eq:LFO}. Magnitudes for DMI are given as $D_{ab}$ and $D_c$ for the nearest-neighbor DM couplings in the $ab$-plane and along the $c$-axis, respectively.\label{Tab:params}}
\end{table}

We treat the low-energy bosonic degrees of freedom in LFO using a Holstein-Primakoff (HP) transformation \cite{HP,doi:10.1146/annurev-conmatphys-031620-104715, Altland, Maestro_2004} which represents the spins in terms of bosonic creation operators $a^\dagger$ and annihilation operators $a$:
\begin{equation}
    S^z = S - a^{\dagger}a,
\end{equation}
\begin{equation}
    S^+ = \sqrt{2S} \sqrt{1-\frac{a^{\dagger}a}{2S}}a \approx \sqrt{2S}a,
\end{equation}
\begin{equation}
    S^- = \sqrt{2S} a^{\dagger} \sqrt{1-\frac{a^{\dagger}a}{2S}} \approx \sqrt{2S}a^{\dagger}.
\end{equation}
where the approximation is the lowest-order term in a Taylor series in $1/S$. The HP bosons are defined relative to a local reference frame in which the $z$-axis aligns with the $z$-component of the spin at that position in the classical ground state. $S^{\alpha}$ and $S^{L \alpha}$ are therefore related by
\begin{equation}
    \vec{S}^L_i = R_i \vec{S}_i,
\end{equation}
where the rotation matrix $R_i$ rotates a vector along the local $z$-axis at position $i$ into the direction of the ground-state spin at $i$.

To find the ground state magnetic configuration, minimization of the classical energy (spins treated as classical variables) is performed via the assumption of a four-sublattice unit cell, due to the small canting introduced by the DMI. Without an applied magnetic field, the classical texture is given by
    $$\vec{S}^L_1 = S(-\cos{\theta}\sin{\phi}, -\cos{\theta}\cos{\phi},\sin{\theta})$$
   $$ \vec{S}^L_2 = S(\cos{\theta}\sin{\phi}, \cos{\theta}\cos{\phi},\sin{\theta})$$
    $$\vec{S}^L_3 = S(-\cos{\theta}\sin{\phi}, \cos{\theta}\cos{\phi},\sin{\theta})$$
\begin{equation}
    \vec{S}^L_4 = S(\cos{\theta}\sin{\phi}, -\cos{\theta}\cos{\phi},\sin{\theta}),
\end{equation}
with $\theta = 0.52$ deg giving a small ferromagnetic canting along the $c$-axis and $\phi = 0.46$ deg giving antiferromagnetic canting within the $ab$-plane \cite{Park_2018}.

\subsection{Dispersion and band Berry curvature}

Expanding around the classical spin configuration above using a $1/S$ expansion, our Hamiltonian takes the form 
\begin{equation} \label{eq:bosonham}
    H \approx \frac{1}{2}\sum_{\vect{r},\vec{\delta}} \psi^{\dagger}(\vect{r}) H_{\vec{\delta}} \psi(\vect{r}+\vec{\delta}),
\end{equation}
where $\vect{r}$ labels the lattice site and $\vec{\delta}$ are the relevant nearest and next-nearest neighbor separation vectors. $\psi(\vect{r})$ is a Nambu spinor, given by
\begin{equation}
    \psi(\vect{r}) = \begin{pmatrix}
        a_1(\vect{r}) \\
        a_2(\vect{r}) \\
        a_3(\vect{r}) \\
        a_4(\vect{r}) \\
        a^{\dagger}_1(\vect{r}) \\
        a^{\dagger}_2(\vect{r}) \\
        a^{\dagger}_3(\vect{r}) \\
        a^{\dagger}_4(\vect{r})
    \end{pmatrix}.
\end{equation}
The non-collinearity of the classical spin configuration will introduce pairing of Holstein-Primakoff magnons (i.e. number non-conserving terms), and thus the Nambu representation is necessary. Here, we have eliminated linear bosonic terms by expanding around the energetic minimum, and we have dropped constants as well as interaction terms of three bosonic operators and higher, consistent with a Taylor expansion in (1/$S$) of the HP transformation.

Performing a Fourier transform as detailed in Appendix \ref{App:BdG} results in the form
\begin{equation} \label{eq:BdG}
    H = \sum_{\vect{k}} \psi^{\dagger}_{\vect{k}} H_{\vect{k}} \psi_{\vect{k}},
\end{equation}
where the $k$-space Nambu spinors are given by 
$$\psi_{\vect{k}} = \begin{pmatrix}
     a_1(\vect{k}) \\
        a_2(\vect{k}) \\
        a_3(\vect{k}) \\
        a_4(\vect{k}) \\
        a^{\dagger}_1(-\vect{k}) \\
        a^{\dagger}_2(-\vect{k}) \\
        a^{\dagger}_3(-\vect{k}) \\
        a^{\dagger}_4(-\vect{k})
\end{pmatrix}.$$

Thus, solving the Fourier transformed problem amounts to diagonalizing the matrix $H_k$. However, care must be taken to ensure that the transformation $T_k$ such that $ T^{\dagger}_k H_k T_k = \Lambda_k$ (with $\Lambda_k$ diagonal) also preserves the bosonic commutation relations
\begin{equation}
    [\psi_k,\psi_{k}^\dagger]=\begin{pmatrix}
                     I_4 & 0 \\ 0 & -I_4
                \end{pmatrix}\equiv\eta,
\end{equation}
so that in the diagonal problem $H = \sum_k \Lambda_k \gamma^{\dagger}_k \gamma_k$ the operators $\gamma^{\dagger}_k \gamma_k$ are number operators. It turns out that for this to be the case, $T_k$ must be a paraunitary transformation \cite{10.1093/ptep/ptaa151,doi:10.1146/annurev-conmatphys-031620-104715, PhysRevB.87.174427,PhysRevB.105.L100402}, which means

 \begin{equation} \label{eq:paraunitary_m}
        T_k \eta T^{\dagger}_k = \eta.
    \end{equation}

\begin{figure}[htp]
    \centering  \includegraphics[width=8.5cm]{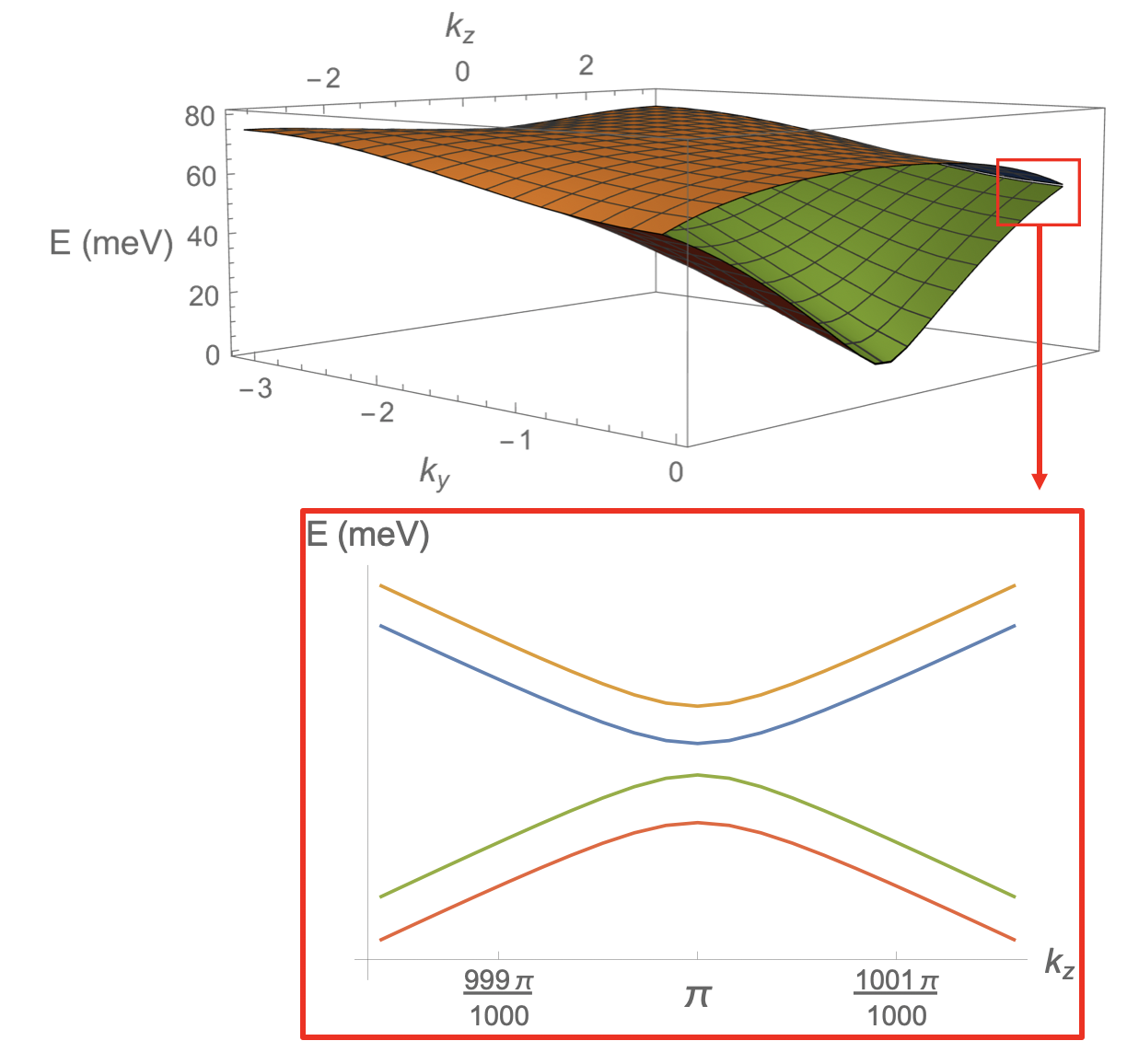}
\caption{\label{Fig:LFOdispersion}TOP: Magnon dispersion $E(0,k_y,k_z)$ for LaFeO$_3$ with an applied field $h = (0,1,1)$. All four particle bands are pictured, with two nearly degenerate upper and two nearly degenerate lower bands. BOTTOM:  Magnon dispersion for LaFeO$_3$ with applied field $h = (0,1,1)$ near $(0,0,\pi)$. A small gap is present, allowing for topological quantities to be computed for each band individually.} 
    \end{figure}

Note that Eq.\eqref{eq:paraunitary_m} implies that rather than diagonalize $H_k$ via similarity transformation in the usual way, the eigenvalue problem we should solve to obtain the magnon dispersion is:
\begin{equation} \label{eq:eigen_m}
        \eta H_k \ket{\psi^n_k} = E^n_k \ket{\psi^n_k}.
    \end{equation}
The ``kets" $\ket{\psi^n_k}$ that solve the eigenproblem are columns of $T_k$, and from paraunitarity they inherit the normalization condition
    \begin{equation} 
        \bra{\psi^m_k} \eta \ket{\psi^n_k} = \eta_{mn}.
    \end{equation}
We call solutions with norm $1$ ``particle bands" and those with norm $-1$ ``hole bands." Equivalently, the bands with positive eigenenergies (norm $1$) are physical particle bands and the bands with opposite sign are the hole partners.  Using this formalism, it is possible to obtain the magnon band structure for LFO, plotted in Fig.~(\ref{Fig:LFOdispersion}). Bulk LFO features pairs of nearly-degenerate bands, as well as near four-fold degeneracies along the Brillouin zone boundary; both of these features make important contributions to the behavior of the SNE.

In addition to the band structure, we can define magnon Berry curvature and Chern numbers. These are defined with respect to the eigenstates in Eq.\eqref{eq:eigen_m}. We first define a Berry connection for the band $n$ as 
 \begin{equation}
        A^n_{\mu} = -i \eta_{nn} \bra{\psi^n_k} \eta \partial_{\mu} \ket{\psi^n_k}.
    \end{equation}
The Berry curvature is then naturally expressed as \cite{Berry_1988},
\begin{equation}
        \Omega^n_{\mu \nu} = \partial_{\mu} A^n_{\nu} - \partial_{\nu} A^n_{\mu}.
    \end{equation}
Finally, the Berry curvature can be used to compute a Chern number by integrating over the first Brillouin zone,
 \begin{equation}
        C^n = \frac{1}{2\pi} \int d^2 k \Omega^n_{xy}. 
    \end{equation}
    
As evident from the formula, which involves a two-dimensional momentum space integral, the Chern number (as well as the spin Nernst response) is an inherently two-dimensional concept. Therefore, in order to discuss it meaningfully, we restrict ourselves to an appropriate two-dimensional problem. Ultimately, we will want to consider the flow of spin in the $z$-direction, with an in-plane temperature gradient which we take to be directed along the $y$-axis.  This geometry naturally invites us to consider a $yz$-plane when computing topological quantities. When computing the total Nernst response, we sum over two-dimensional contributions from all planes. That is, if $\alpha^{\mu}_{\nu \beta}(k_x)$ is the contribution to the SNE from the $k_x$ plane, then the total response will be given by $\alpha^{\mu}_{\nu \beta} = \int dk_x \; \alpha^{\mu}_{\nu \beta}(k_x)$ \cite{3DHall}.

\begin{figure}[htp]
    \centering
\includegraphics[width=8.5cm]{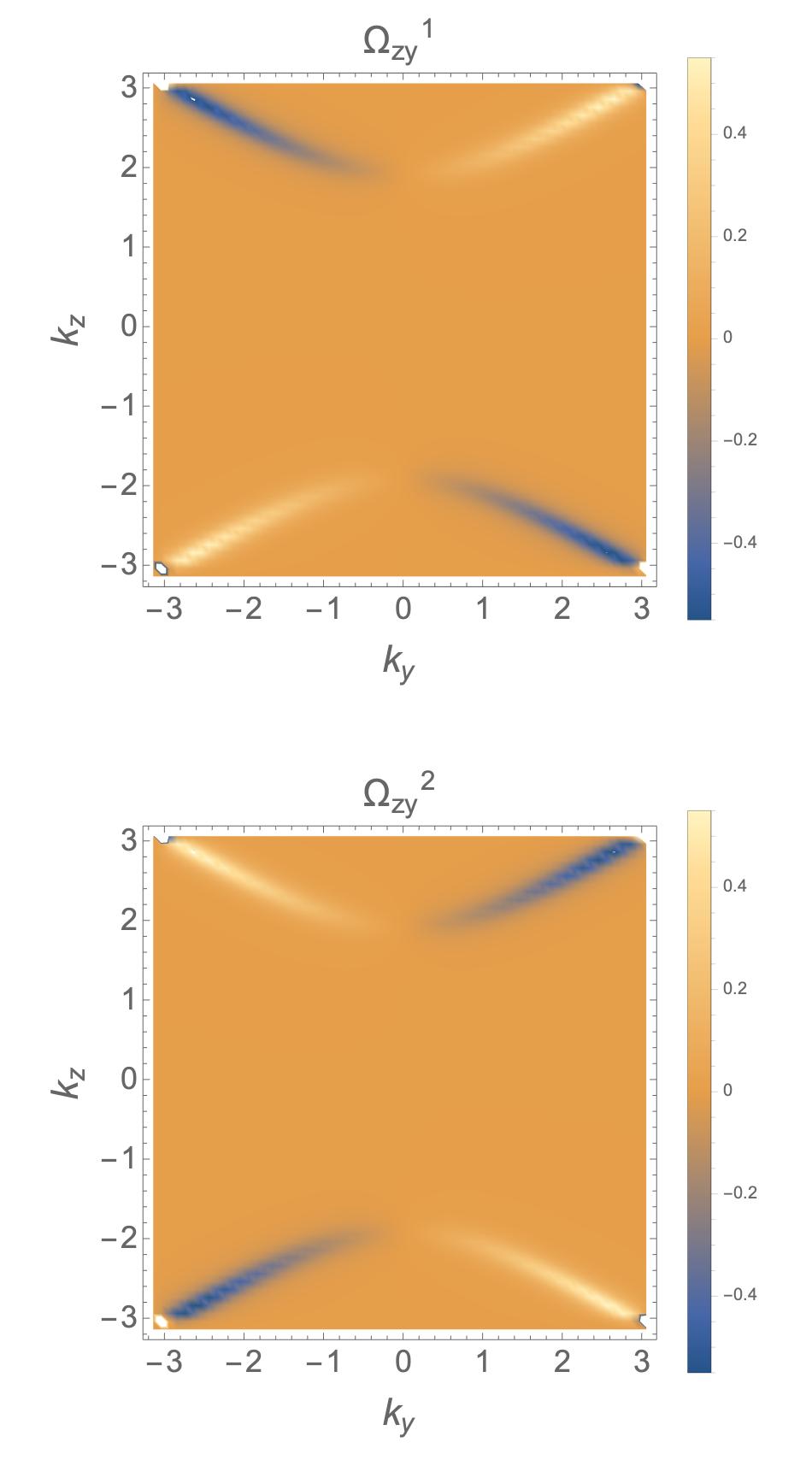}   \caption{\label{Fig:LFOberryfree_m}Magnon Berry curvature $\Omega^n_{zy}(0,k_y,k_z)$ for LaFeO$_3$ with applied field $h = (0,1,1)$, for $n=1,2$. The lower bands $n=3,4$ have mostly trivial Berry curvature, with hotspots only at corners, shown in App. \ref{App:Berry}.} 
    \end{figure}

Finally, we note that $\Omega^n$ can be represented in the Thouless-Kohmoto-Nightingale-Nijs (TKNN) \cite{TKNN} form derived from linear response theory, analogous to the one we will present for the spin Berry curvature,
 \begin{equation} \label{eq:bandberry}
        \Omega^{n}_{\mu \nu}(k) = \sum_{m \neq n} \eta_{nn}\eta_{mm}\frac{2 {\rm Im}\bra{\psi^n_k} \partial_{\mu}H_k\ket{\psi^m_k}\bra{\psi^n_k} \partial_{\nu} H_k \ket{\psi^m_k}}{(E_{nk}-E_{mk})^2},
    \end{equation}
where the sum is taken over both particle and hole bands. The quantity $\Omega^n_{zy}(k_x = 0)$ for each particle band is plotted in Fig.(\ref{Fig:LFOberryfree_m}). We see that each band is topologically trivial ($C^n = 0$) because when integrating over the BZ each point of positive curvature is cancelled by another of negative curvature; this is from the symmetries $\Omega^n_{zy}(k_x,k_y,k_z) = -\Omega^n_{zy}(-k_x,-k_y,k_z)$ and $\Omega^n_{zy}(k_x,k_y,k_z) = -\Omega^n_{zy}(k_x,k_y,-k_z)$, either of which alone is enough to enforce $C^n = 0$. 
%\textcolor{blue}{(BM: Is there a simple physical picture for this symmetry? For example, mirror symmetry?)}

Next, we look at the SNE itself. The spin-current continuity equation is written as $\frac{\partial \vec{S}}{\partial t} = -\nabla \cdot \vec{j}_S + \vec{\tau}_S$ \cite{PhysRevResearch.2.013079,PhysRevB.89.054420, Nakata2011-dp,PhysRevLett.96.076604}. In an inversion-symmetric system, the torque term $\vec{\tau}_S$ does not contribute to the SNE. The response tensor $\alpha$ is given by \cite{PhysRevResearch.2.013079,PhysRevB.89.054420},
\begin{equation} \label{eq:SNE}
  J^{\gamma}_{\lambda} =  \alpha^{\gamma}_{\lambda \beta}\nabla_{\beta}T = \frac{2 k_B}{V} \sum_n \sum_{k} \Omega^{n\gamma}_{\lambda \beta} c_1[g(E_{nk})]\nabla_{\beta}T,
    \end{equation}
where $g(E)$ is the Bose occupation factor and $c_1(x) = (1+x)\ln(x) -x\ln(x)$. The sum $\sum_n$ is over only particle bands and the sum $\sum_k$ is taken over all three dimensions. Here $\gamma$ is the polarization of the spin current, $\lambda$ is the direction of the spin current, and $\beta$ is the direction of the temperature gradient (each of which could be along $x,y$ or $z$). The quantity $\Omega^{n\gamma}_{\lambda \beta}$ is the spin Berry curvature, a generalization of the ``band" Berry curvature $\Omega^n_{\mu \nu}$. Analogous to Eq.\eqref{eq:bandberry}, it is given by \cite{PhysRevLett.117.217203, PhysRevB.89.054420},
\begin{equation} \label{eq:spinberry}
        \Omega^{n\alpha}_{\beta \gamma}(k) = \sum_{m \neq n} \eta_{nn}\eta_{mm}\frac{2 {\rm Im}\bra{\psi^n_k} j^{\alpha}_{\beta}\ket{\psi^m_k}\bra{\psi^n_k} \partial_{\gamma} H_k \ket{\psi^m_k}}{(E_{nk}-E_{mk})^2},
    \end{equation}
where $j^{\alpha}_{\beta}$ is the spin current operator in the BdG representation. The spin current operator is given by,
\begin{equation}
    j^{\alpha}_{\beta} = \frac{1}{4} (\partial_{\beta} H_k \eta \Sigma^{\alpha} + \Sigma^{\alpha} \eta \partial_{\beta} H_k),
    \end{equation}
with $\Sigma^{\alpha} = {\rm diag} (S^\alpha_1,S^\alpha_2,S^\alpha_3,S^\alpha_4,S^\alpha_1,S^\alpha_2,S^\alpha_3,S^\alpha_4)$ encoding the magnon polarization, for $S^\alpha_i$ the classical spin configuration. Fig.(\ref{Fig:LFOspinberryfree}) depicts the spin Berry curvature,  $\Omega^{nx}_{zy}(0,k_y,k_z)$, for each band.

\begin{figure}[htp]
    \centering  \includegraphics[width=8.5cm]{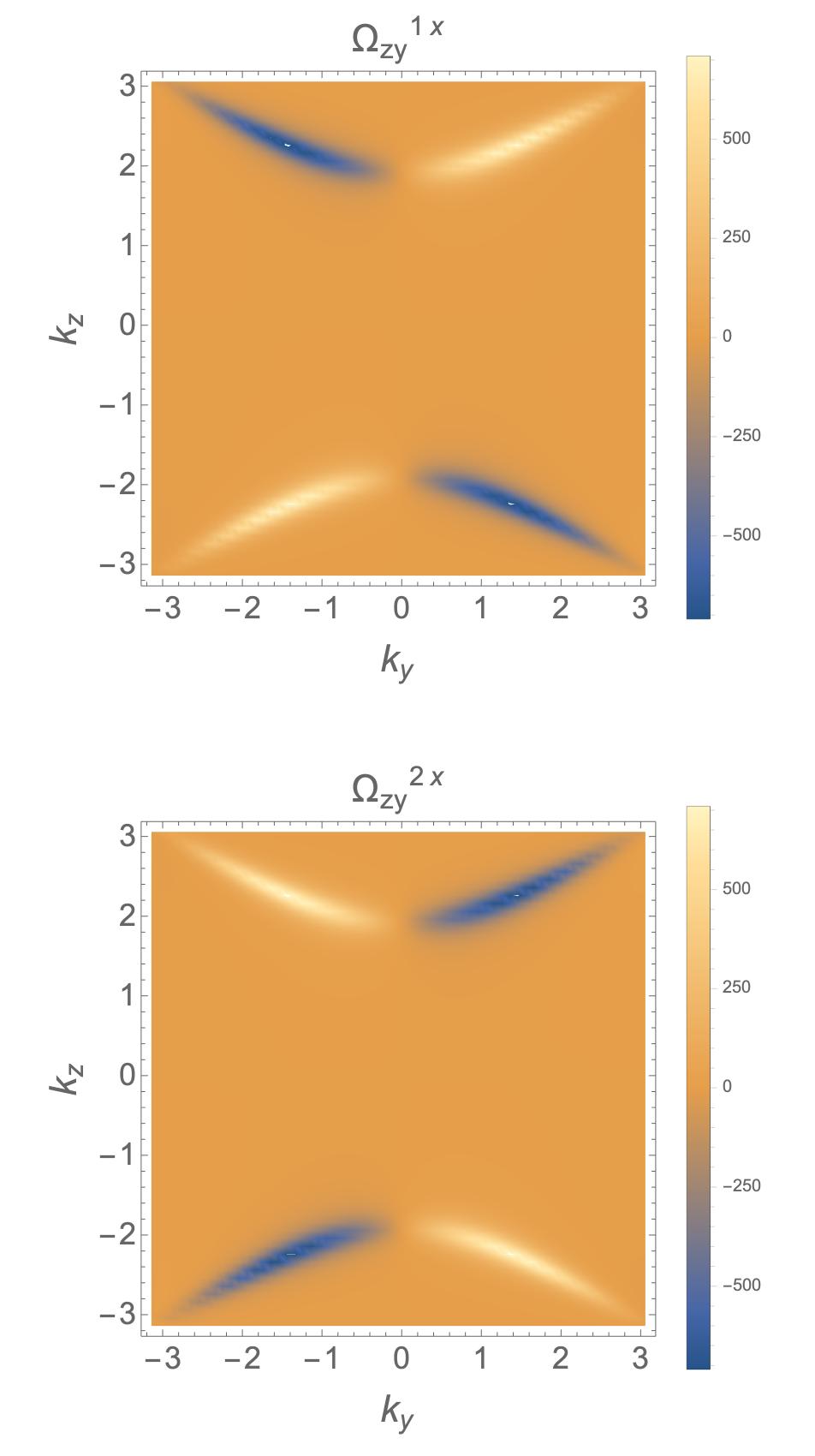}   \caption{\label{Fig:LFOspinberryfree}Magnon spin Berry curvature $\Omega^{nx}_{zy}(0,k_y,k_z)$ for LaFeO$_3$ with applied field $h = (0,1,1)$, for $n=1,2$, which make the dominant contributions to the spin Berry curvature. Plots for all bands are found in App.\ref{App:Berry}.} 
    \end{figure}
We find the symmetries $$\Omega^{nx}_{zy}(k_x,k_y,k_z) = -\Omega^{nx}_{zy}(-k_x,-k_y,k_z),$$ and $$\Omega^{nx}_{zy}(k_x,k_y,k_z) = -\Omega^{nx}_{zy}(k_x,k_y,-k_z),$$ which taken together with the corresponding symmetries of the bands $E_n(k_i) = E_n(-k_i)$ (in other words, changing the sign of any $k_i$ does not change the energy) lead to a vanishing SNE within each band. %Furthermore, even if this inversion symmetry were broken there would be near-cancellation between bands, due to the intraband relations $\Omega^{1(3)x}_{zy}(0,k_y,k_z) = -\Omega^{2(4)x}_{zy}(0,k_y,k_z)$ and the near-degeneracies $E_{1(3)} \approx E_{2(4)}$. 
Whereas the symmetries of the dispersion did not matter in the cancellation of the Chern number, they are important here because they enter through the Bose occupation function $g$.

Now that we have discussed the relevant features of unperturbed bulk LFO, we turn next to an analysis of what sorts of modifications to the model would create a nonzero SNE. A weak dimerization will turn out to play an important role in producing a finite SNE.

\section{Symmetry breaking and spin Nernst effect} \label{Sec:symmetry}

We begin by asking the general question: at the level of the BdG Hamiltonian $H_k$, what symmetries would need to be lowered in order to produce a finite SNE?  We have already begun to discuss this above when we commented on the elimination of both the Chern numbers and the SNE within each band as well as between bands - the intra-band cancellation occurs due to $k$-space symmetries in the Berry curvature.

%The second contributor greatly reduces but does not exactly eliminate the SNE in bulk LFO, even when the above inversion symmetry is broken. This is due to an approximate inter-band symmetry: $$(\Omega^s_{n\vec{k}})^{\gamma}_{yz} = -(\Omega^s_{m\vec{k}})^{\gamma}_{yz}$$
%where $m$ and $n$ are nearly degenerate bands $E_{n\vec{k}}\approx E_{m\vec{k}}$, and therefore have a similar Bose occupation factor. This leads to opposite contributions to $\alpha,$ which results in a very small SNE even when the intra-band symmetry is broken.

\subsection{$k$-space symmetry}

The intraband symmetry that leads to the vanishing of the SNE manifests at the level of the BdG Hamiltonian as 
\begin{equation}
    H(k_x,k_y,k_z) = H(k_x,k_y,-k_z),
\end{equation}
\begin{equation}
    H(k_x,k_y,k_z) = H(k_x,-k_y,k_z),
\end{equation}
where for clarity we have switched to notation where $H_{\vect{k}} = H(k_x,k_y,k_z)$. We note at the outset that not only will breaking these symmetries create a net SNE via Eq.\eqref{eq:SNE}, but can also create a source term contribution due to $H_{\vect{k}} \neq H_{-\vect{k}}$. %{\color{red} is this obvious?, important?}. 
Below we focus on the contribution from the current term rather than the source term. The presence of a source term due to inversion-symmetry breaking will lead to some dissipation; however, all of our results will apply within the window of the spin relaxation time \cite{PhysRevLett.96.076604}. 

We first consider the $k_z \to -k_z$ symmetry. For a term to break this symmetry, it must appear in the Hamiltonian in the form 
\begin{equation} \label{eq:ansatz}
    [H_{\Delta}(k_z)]_{ij} = \Delta_{ij} \sin(k_z l_{ij} ),
\end{equation}
(or in some other odd function of $k_z$)
where we are so far not restricting the elements of this term in particle-hole space, but are simply enforcing that the whole term is odd in $k_z$. Since any term with the property $f(k) \neq f(-k)$ can be written in terms of its Fourier components, terms of the form Eq.\eqref{eq:ansatz} with various $l_{ij}$ are sufficient to specify any inversion-breaking term. We can therefore simply ask what sort of term at the level of the spin Hamiltonian gives rise to a bosonic term of the form Eq.\eqref{eq:ansatz}.

If we restrict ourselves only to nearest-neighbor couplings that break inversion symmetry, then it turns out that the most general form of the symmetry-breaking term is (see Appendix \ref{App:HD}),
\begin{equation} \label{eq:hdelta}
 H_{\Delta} = \frac{1}{4} \sum_j \Gamma^{\alpha \beta} (S^{\alpha}_{aj}S^{\beta}_{bj-1} - S^{\alpha}_{aj}S^{\beta}_{bj}),
\end{equation}
 where we do not enforce a specific form for $\Gamma^{\alpha \beta}$, except that it be real, which is required for $H_{\Delta}$ to be Hermitian. The sum on $j$ is taken along the axis on which the symmetry is to be broken. Rotating into the local frame such that $\tilde{\Gamma}^{\alpha \beta} = \Gamma^{\gamma \delta} R^{\gamma \alpha}_a R^{\delta \beta}_b$ and performing the Holstein-Primakov transformation and Fourier transformations we find that,
\begin{widetext}
\begin{align*}
    H^{x,y}_{\Delta} = \frac{1}{4}\sum_d \sum_k (-1)^d [&(\tilde{\Gamma}^{xx} - \tilde{\Gamma}^{yy}-i \tilde{\Gamma}^{xy}-i \tilde{\Gamma}^{yx})e^{-ik(d+x_b)}a_k b_{-k} 
    \\
+&(\tilde{\Gamma}^{xx} + \tilde{\Gamma}^{yy}+i \tilde{\Gamma}^{xy}-i \tilde{\Gamma}^{yx})e^{-ik(d+x_b)}a_k b^{\dagger}_{k} 
   \\ - &(\tilde{\Gamma}^{xx} + \tilde{\Gamma}^{yy}-i \tilde{\Gamma}^{xy}+i \tilde{\Gamma}^{yx})e^{ik(d+x_b)}a^{\dagger}_k b_{k} \\
    -&(\tilde{\Gamma}^{xx} - \tilde{\Gamma}^{yy}+i \tilde{\Gamma}^{xy}+i \tilde{\Gamma}^{yx})e^{ik(d+x_b)}a^{\dagger}_k b^{\dagger}_{-k}],
\end{align*}
\end{widetext}
where $H^{x,y}_{\Delta}$ denotes that we are only considering the $(x, y)$ elements of $\tilde{\Gamma}$. Here $\sum_d$ is over the nearest-neighbor displacements $d = 0,-1$ and $x_b = 1/2$ is the separation between $a$ and $b$ sublattices. We see that when we carry out the $d$ sum, the factor $(-1)^d$ is what creates odd $k$-dependence. This factor comes in particular from the relative minus sign between coupling of $S^{\alpha}_{aj}$ to $S^{\beta}_{bj}$ and $S^{\beta}_{bj-1}$. Performing this sum we explicitly find,
\begin{align*}
    H^{x,y}_{\Delta} = \frac{-i}{2} \sum_k  \sin(k/2) [&(\tilde{\Gamma}^{xx} - \tilde{\Gamma}^{yy}-i \tilde{\Gamma}^{xy}-i \tilde{\Gamma}^{yx})a_k b_{-k} 
    \\
+&(\tilde{\Gamma}^{xx} + \tilde{\Gamma}^{yy}+i \tilde{\Gamma}^{xy}-i \tilde{\Gamma}^{yx})a_k b^{\dagger}_{k} 
   \\ - &(\tilde{\Gamma}^{xx} + \tilde{\Gamma}^{yy}-i \tilde{\Gamma}^{xy}+i \tilde{\Gamma}^{yx})a^{\dagger}_k b_{k} 
    \\-&(\tilde{\Gamma}^{xx} - \tilde{\Gamma}^{yy}+i \tilde{\Gamma}^{xy}+i \tilde{\Gamma}^{yx})a^{\dagger}_k b^{\dagger}_{-k}].
\end{align*}
 Note that in the presence of an inversion-breaking term in the BdG Hamiltonian is not sensitive to the particular form of $\Gamma$, but rather depends on the presence of opposite coupling between ``forward" and ``backward" nearest neighbors. 

We have so far found a general form for spin coupling along some direction which  introduces $k$-space inversion-breaking terms along that direction in the BdG Hamiltonian. It is useful to note a special feature of this term, which happens to make its analysis simpler. In principle, when one adds $H_{\Delta}$ to the original Hamiltonian, one should minimize the classical energy once again to find the new classical ground state. However, for the form we have proposed, if one restrict themselves to the four-sublattice problem as we have for unperturbed LFO, $H_{\Delta}$ turns out not to modify the classical energy, and therefore does not modify the canting. This is because in the classical limit $H_{\Delta}$ becomes, for a uniform spin configuration,
\begin{equation}
    H_{\Delta} = \frac{1}{4} \sum_j \tilde{\Gamma}^{\alpha \beta}(S^{\alpha}_{a} S^{\beta}_{b} - S^{\alpha}_{a} S^{\beta}_{b}) = 0,
\end{equation}
for any arrangement where each sublattice is uniform from site to site - for example, where there is not a spiral texture along a sublattice.

To check that it is sensible to proceed with this uniform-sublattice assumption, one can check that $H_{\Delta}$ does not introduce linear boson operators when one performs the HP expansion: if it were to introduce linear terms, this would indicate that one were not expanding around an energy minimum anymore. To see that linear terms are not introduced for any choice of rotations, we let $i = x,y$ and write out the only terms that are able to contribute linear terms, which are those involving $S^i_j S^z_l$. The terms in $H_{\Delta}$ that could generate linear $a$ bosons are 
\begin{dmath}
    \sum_j \tilde{\Gamma}^{iz} (S^i_{aj} S^z_{bj-1} - S^i_{aj} S^z_{bj}) = \sum_j \tilde{\Gamma}^{iz} S^i_{aj}(S^z_{bj-1} -S^z_{bj}).
\end{dmath}
When one performs the HP transformation on $S^z$, the $\mathcal{O}(S)$ terms that would normally contribute a linear term cancel one another:
\begin{dmath}
    \sum_j \tilde{\Gamma}^{iz} S^i_{aj}(S - b^{\dagger}_{j-1}b_{j-1} - S +b^{\dagger}_jb_{j}) = \sum_j \tilde{\Gamma}^{iz} S^i_{aj}(- b^{\dagger}_{j-1}b_{j-1} +b^{\dagger}_jb_{j}).
\end{dmath}
The only contributions from these kinds of terms are boson interactions, which we drop in the $1/S$ expansion. A similar argument holds if we look for the corresponding terms that would generate linear $b$ bosonic terms. Therefore, we find that: 
\begin{enumerate}
    \item $H_{\Delta}$ does not modify the classical ground state configuration. 
    \item The lowest-order bosonic terms $H_{\Delta}$ contributing to the Hamiltonian are quadratic.
\end{enumerate}
Therefore, we can safely proceed with adding terms of the form $H_{\Delta}$ to the Hamiltonian, without needing to find a new energetic minimum configuration. In fact, no $z$-spin components contribute nonzero boson terms except at higher than quadratic order. Therefore, the expression we wrote for $H^{x,y}_{\Delta}$ is already the full expression for $H_{\Delta}$ to quadratic order. Expressing it in BdG-doubled form, we have:
\begin{widetext}
\begin{align} \label{eq:c}
    H^c_{\Delta} = \frac{-i}{4} \sum_k  \sin(k/2) [&(\tilde{\Gamma}^{xx} - \tilde{\Gamma}^{yy}-i \tilde{\Gamma}^{xy}-i \tilde{\Gamma}^{yx})(b_{-k} a_{k}  - a_{-k} b_k ) \nonumber
    \\
+&(\tilde{\Gamma}^{xx} + \tilde{\Gamma}^{yy}+i \tilde{\Gamma}^{xy}-i \tilde{\Gamma}^{yx})(b^{\dagger}_{k} a_k  -a_{-k} b^{\dagger}_{-k}) \nonumber
   \\ - &(\tilde{\Gamma}^{xx} + \tilde{\Gamma}^{yy}-i \tilde{\Gamma}^{xy}+i \tilde{\Gamma}^{yx})(a^{\dagger}_k b_{k} -b_{-k} a^{\dagger}_{-k} ) \nonumber
    \\-&(\tilde{\Gamma}^{xx} - \tilde{\Gamma}^{yy}+i \tilde{\Gamma}^{xy}+i \tilde{\Gamma}^{yx})(a^{\dagger}_k b^{\dagger}_{-k}-b^{\dagger}_k a^{\dagger}_{-k})].
\end{align}
\end{widetext}

The $c$ superscript denotes that this is valid for couplings along the $c$-axis, because so far we have assumed that the coupled spins are arranged along a line parallel to the axis along which the inversion symmetry is being broken. This is true of couplings between $S_1$ and $S_2$ as well as between $S_3$ and $S_4$, which lie along the $c$-axis and break $k_z$-inversion. However, couplings along the $b$-axis which break the $k_y$ symmetry are slightly more complicated due to the fact that nearest neighbors are not separated by just $\hat{y}/2$, but instead by $\frac{\pm\hat{x}+\hat{y}}{2}$. The result  is that couplings in the $ab$-plane take the form,
\begin{widetext}
\begin{align} \label{eq:ab}
    H^{ab}_{\Delta} = \frac{-i}{2} \sum_k \cos{(k_x/2)} \sin(k_y/2) [&(\tilde{\Gamma}^{xx} - \tilde{\Gamma}^{yy}-i \tilde{\Gamma}^{xy}-i \tilde{\Gamma}^{yx})(b_{-k} a_{k}  - a_{-k} b_k ) \nonumber
    \\
+&(\tilde{\Gamma}^{xx} + \tilde{\Gamma}^{yy}+i \tilde{\Gamma}^{xy}-i \tilde{\Gamma}^{yx})(b^{\dagger}_{k} a_k  -a_{-k} b^{\dagger}_{-k}) \nonumber
   \\ - &(\tilde{\Gamma}^{xx} + \tilde{\Gamma}^{yy}-i \tilde{\Gamma}^{xy}+i \tilde{\Gamma}^{yx})(a^{\dagger}_k b_{k} -b_{-k} a^{\dagger}_{-k} ) \nonumber
    \\-&(\tilde{\Gamma}^{xx} - \tilde{\Gamma}^{yy}+i \tilde{\Gamma}^{xy}+i \tilde{\Gamma}^{yx})(a^{\dagger}_k b^{\dagger}_{-k}-b^{\dagger}_k a^{\dagger}_{-k})].
\end{align}
\end{widetext}

Including the terms in Eq.\eqref{eq:c} and Eq.\eqref{eq:ab} the Hamiltonian has the effect of creating a nonzero contribution to the SNE from each individual band. However, for large enough perturbations there is not only a net contribution from each band, but a total contribution when summed over bands, resulting in a finite SNE. This is shown explicitly below.

\subsection{Physical model for symmetry breaking}
%\textcolor{blue}{(BM: Is it possible to directly begin from Eq. (36) from any physical insights? If so, could it be better to move many derivation details in part A into the appendix?)}
For concreteness, we consider a model with exchange-like symmetry-breaking coupling of the form
\begin{equation}
     H_{\Delta} = \sum_j \Delta (S^{\alpha}_{aj}S^{\alpha}_{bj-1} - S^{\alpha}_{aj}S^{\alpha}_{bj}).
\end{equation}
We begin by noting that a possible origin of a term like this can be considered by absorbing $H_{\Delta}$ into the relevant exchange term of a magnetic Hamiltonian, $H_J = J\sum_j (S^{\alpha}_{aj} S^{\alpha}_{bj} + S^{\alpha}_{aj} S^{\alpha}_{bj-1})$. Collecting terms, we see that
\begin{dmath}
    H_{J} + H_{\Delta} =  \sum_j ((J+\Delta)S^{\alpha}_{aj}S^{\alpha}_{bj-1} +(J-\Delta)S^{\alpha}_{aj}S^{\alpha}_{bj}).
\end{dmath}
Thus, the effect of the symmetry-breaking term $H_{\Delta}$ is to create a difference $\Delta$ between the exchange coupling for atoms within the same unit cell and those in different unit cells. This would occur, for example, in the case of dimerization along the axis of symmetry-breaking.

In LFO, we require symmetry-breaking along the $y$- and $z$-axes. We therefore require the full inversion-breaking term to be
\begin{dmath} \label{eq:uniformpert}
     H_{\Delta} = \sum_r \left[\Delta (S^{\alpha}_{1}(\vec{r})S^{\alpha}_{2}(\vec{r}-\hat{e}_c) - S^{\alpha}_{1}(\vec{r})S^{\alpha}_{2}(\vec{r})) + \Delta (S^{\alpha}_{4}(\vec{r})S^{\alpha}_{3}(\vec{r}-\hat{e}_c) - S^{\alpha}_{4}(\vec{r})S^{\alpha}_{3}(\vec{r})) + \Delta (S^{\alpha}_{1}(\vec{r})S^{\alpha}_{4}(\vec{r}-\hat{e}_b) - S^{\alpha}_{1}(\vec{r})S^{\alpha}_{4}(\vec{r})) + \Delta (S^{\alpha}_{1}(\vec{r})S^{\alpha}_{4}(\vec{r}-\hat{e}_b - \hat{e}_a) - S^{\alpha}_{1}(\vec{r})S^{\alpha}_{4}(\vec{r}-\hat{e}_a)) + \Delta (S^{\alpha}_{2}(\vec{r})S^{\alpha}_{3}(\vec{r}-\hat{e}_b) - S^{\alpha}_{2}(\vec{r})S^{\alpha}_{3}(\vec{r})) + \Delta (S^{\alpha}_{2}(\vec{r})S^{\alpha}_{3}(\vec{r}-\hat{e}_b - \hat{e}_a) - S^{\alpha}_{2}(\vec{r})S^{\alpha}_{3}(\vec{r}-\hat{e}_a)) \right].
\end{dmath}
Couplings along the $c$-axis produce magnon terms of the form $H^c_{\Delta}$ [Eq.\eqref{eq:c}], while couplings in the $ab$-plane (which break the $k_y$ inversion symmetry) produce terms of the form $H^{ab}_{\Delta}$ [Eq.\eqref{eq:ab}]. We can now ask about what changes occur to the band structure, Berry curvature, Chern numbers, and SNE when Eq.\eqref{eq:uniformpert} is added to the Hamiltonian. 

%The band structure is largely unchanged by Eq. \eqref{eq:uniformpert}, with differences manifesting near the edges of the BZ as depicted in Fig

%\begin{figure}[htp]
    %\centering
    %\includegraphics[width=8.5cm]{pertdispersion.png}
    
   %\caption{\label{Fig:pertdispersion}Difference between old and new dispersion for highest band, where new dispersion is for $\Delta = 0.5$, or $\Delta/J \sim 0.1$. We see that the perturbation to the band structure occurs mostly at the BZ boundary.}
   % \end{figure}

%Using finite-size scaling to overcome difficulties in numerically computing $\alpha$ for large system sizes, we can see that for a small perturbation \eqref{eq:uniformpert} the net spin Nernst effect is still zero across bands. This is depicted in Fig.(\ref{Fig:nernstscaling}).

%\begin{figure}[htp]
   % \centering
    %\includegraphics[width=8.5cm]{nernstscaling.jpeg}
    
  % \caption{\label{Fig:nernstscaling}The spin Nernst coefficient $\alpha^x_{zy}$ for spin polarization along $x$, injected along $z$, for temperature gradient along $y$, plotted against inverse linear system size, for a square system of $A = L_y L_z$ in the $yz$-plane. The data corresponds to $\Delta = 0.01$ and $k_B T = 100$. We see that the relationship becomes linear for large system sizes, and the coefficient converges to 0. }
   % \end{figure}

For small perturbation strengths we recover a nonzero bandwise SNE; however, while the contributions from individual bands are nonzero, as is expected because of the lower symmetry of $\Omega^x_{zy}$, the effects from nearly-degenerate bands cancel one another, leading to a vanishing total SNE. For example, at $\Delta = 0.01$ meV and $k_B T = 100$ meV, we find that for the highest band $|\alpha^{1x}_{zy}|/k_BT = 0.033$, where $\alpha^{x}_{zy} = \sum_{n} \alpha^{nx}_{zy}$ with the sum taken over particle bands, but $\alpha^x_{zy} = 0$.

However, when we increase the perturbation size we find that large spin Berry curvature hotspots are created, which are large enough to overcome the effect of near-degenerate bands with opposite $\Omega^{nx}_{zy}$. When $\Omega^{nx}_{zy}$ is small, the fact that the two highest and two lowest bands are nearly degenerate leads to cancellation of the SNE in Eq.\eqref{eq:SNE}; when $\Omega^{nx}_{zy}$ becomes large in some region, even a small gap leading to a small difference in Bose occupation factor is enough to obtain a net effect. For example, we find that for $\Delta/J = 0.1$ the SNE increases to $\alpha^{x}_{zy}/k_B T = 2.5 \times 10^{-4}$, which is comparable in magnitude to similar studies in 2D and 3D materials \cite{3DHall}. 

\begin{figure}[htp]
    \centering
\includegraphics[width=8.5cm]{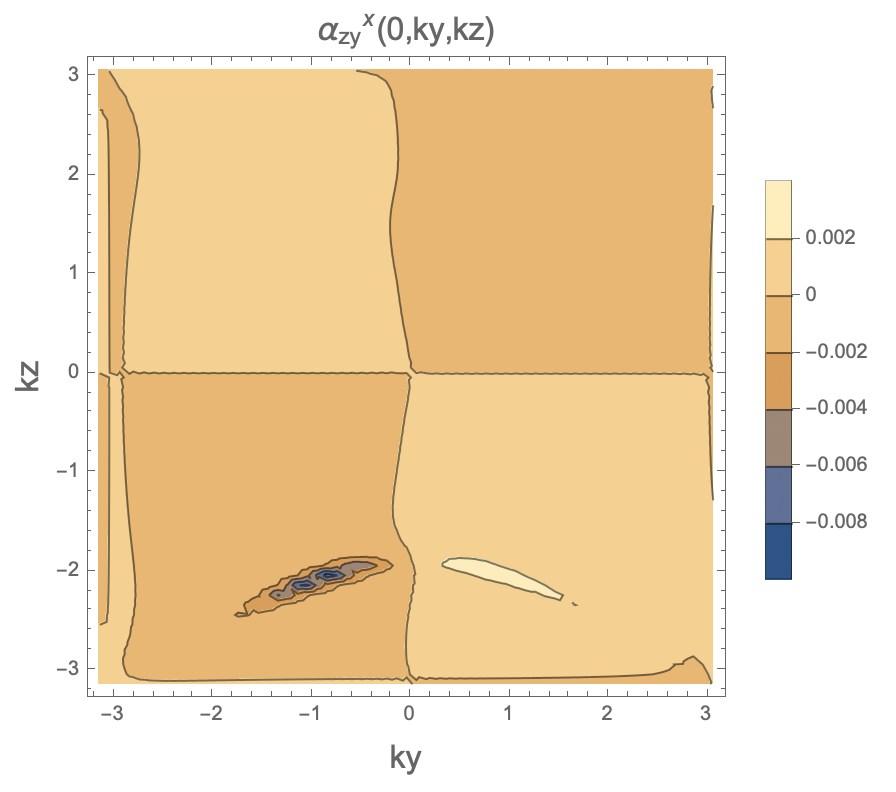}
\caption{\label{Fig:nernstcoeff} The total $k$-space contribution to the SNE. When the spin Berry curvature becomes large through symmetry breaking, an asymmetrical  hotspot  in contribution arises, which is enough for a finite net SNE.}
    \end{figure}

Fig.~\ref{Fig:nernstcoeff} shows $\alpha^x_{zy}(0,k_y,k_z),$ the $k$-space contribution to the spin Nernst coefficient such that $\alpha^x_{zy}(k_x = 0) = \int d^2k \; \alpha^x_{zy}(0,k_y,k_z)$. Here we clearly see that the net effect is due to an asymmetrical feature in $\alpha^x_{zy}(0,k_y,k_z)$ which is not cancelled by contributions elsewhere in the BZ. Furthermore, such an asymmetrical feature exists for general $k_x$ contributions $\alpha^x_{zy}(k_x)$, and not only at $\alpha^x_{zy}(0)$. Since our system obeys $H(k_x) \neq H(-k_x)$, these finite contributions across $k_x$ values do not cancel one another, leading to a finite total SNE $\alpha^x_{zy} = \int dk_x \; \alpha^x_{zy}(k_x)$.

\begin{figure}[htp]
\includegraphics[width=8.5cm]{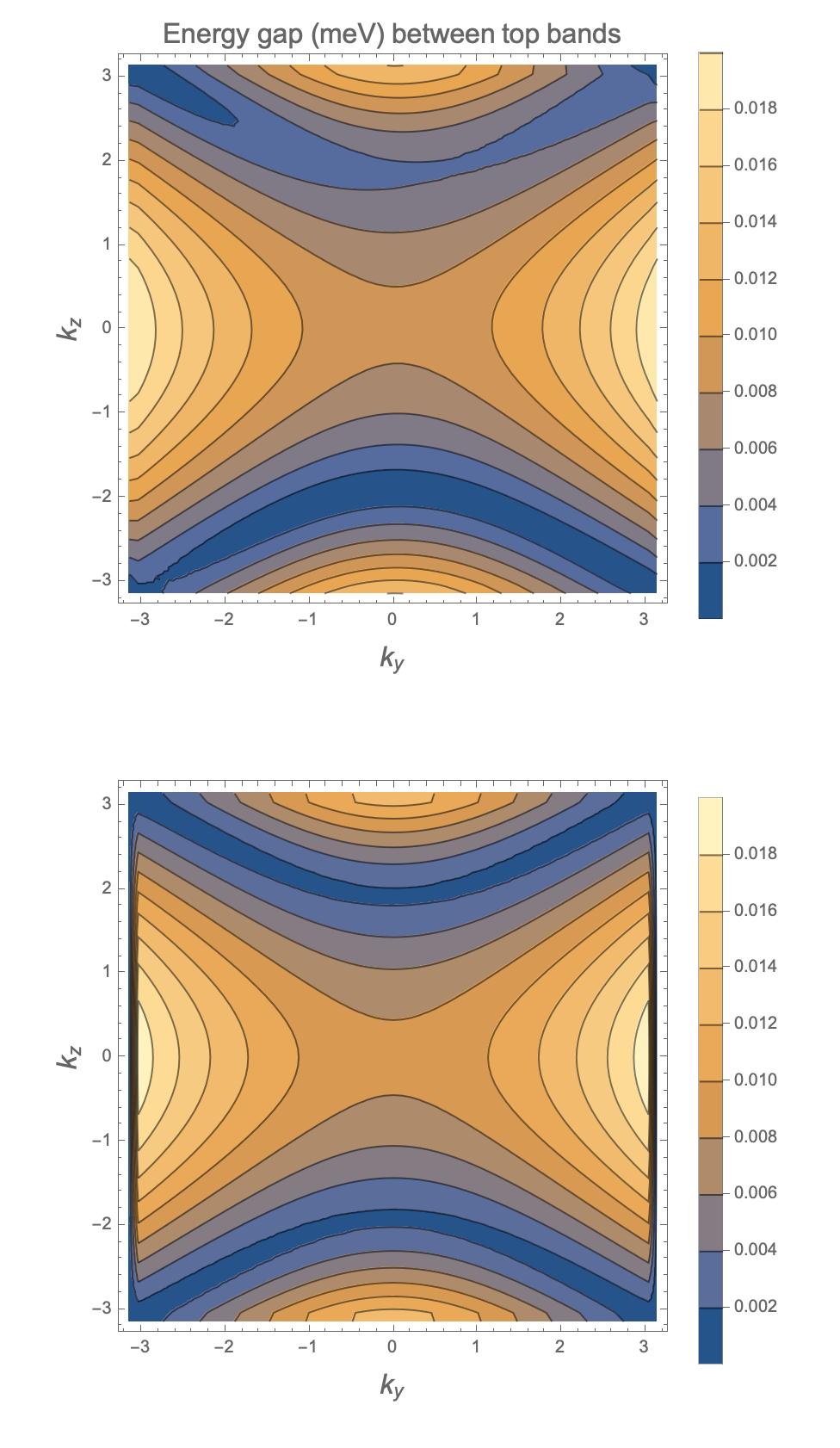}   \caption{\label{Fig:energydiff} Energy gap between two highest bands. (TOP) The difference between the energies of the top two bands after symmetry breaking. (BOTTOM) The same quantity computed before symmetry breaking.}
    \end{figure}

The asymmetrical contribution to the SNE from the locus of points in the third quadrant of the BZ shown in Fig.~\ref{Fig:nernstcoeff} arise from changes to the band structure induced by symmetry breaking. Most importantly, a curve of near-degeneracy between the highest two bands imparts the large spin Berry curvature required to produce a SNE, and the asymmetrical contribution of that spin Berry curvature arises due to the symmetry breaking. This is shown in the upper panel in Fig.\ref{Fig:energydiff}. By comparison of Fig.~\ref{Fig:nernstcoeff} with Fig.~\ref{Fig:energydiff}, one can see the origin of the locus of points with largest spin Berry curvature comes from the regions with the smallest gap. The symmetry of this locus of points is lowered when symmetry is broken.

\begin{figure}[htp] \includegraphics[width=8.5cm]{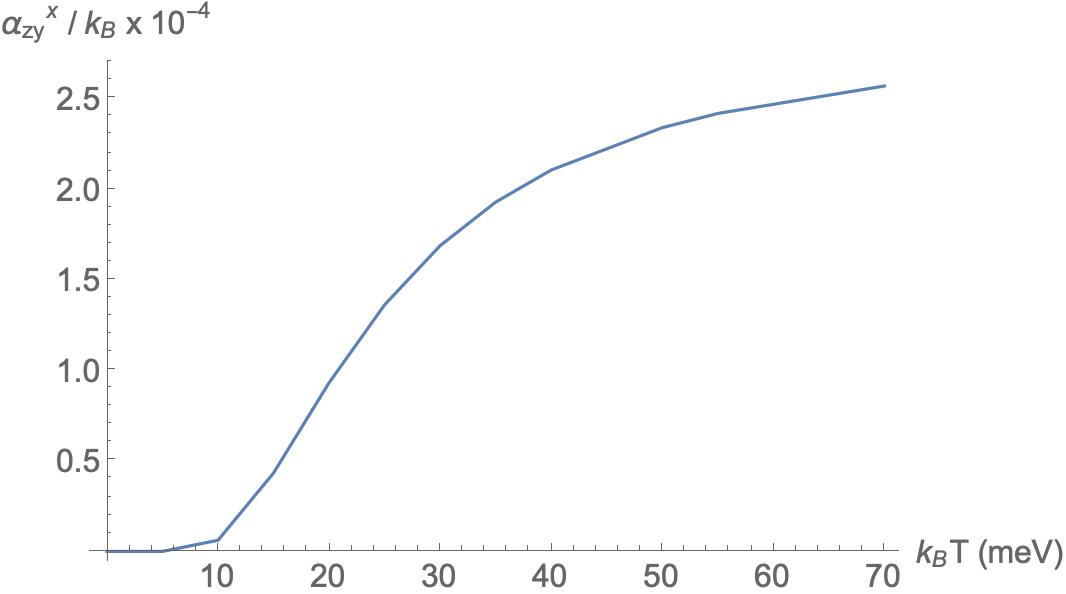}
\caption{\label{Fig:tempdepend} SNE as a function of temperature, for $\Delta/J$ = 0.1.}
    \end{figure}

Finally, we present the temperature dependence of the SNE in the presence of symmetry breaking in Fig.\ref{Fig:tempdepend}. Because the major contributions to the SNE come from the higher-energy bands, we see that higher temperatures lead to a greater SNE due to larger occupation of these bands. It is also important to note that for LFO, $T_N k_B \approx 80$ meV, so this symmetry breaking should lead to an observable SNE even below the critical point.

\section{Conclusion} \label{Sec:conclusion}
In this work we have shown that while unstrained/undimerized LFO does not support a SNE that would explain the results of Lin et al.\cite{LFOExp}, under a particular lowering of the bulk symmetry the SNE can become nonzero. This can occur through physical mechanisms such as dimerization along the $y$- and $z$-axes. This effect could explain the anomalous spin transport between LFO and Pt/W observed by Lin et al. \cite{LFOExp}. 

We note that our work also serves a more general purpose for the analysis of transport phenomena in magnetic systems. We have introduced a framework for relaxing cancellation effects at the level of generalized Berry curvature which usually result in the vanishing of nontrivial currents. In our case, this cancellation was due to inversion symmetries in the spin Berry curvature. We systematically analyzed the necessary and sufficient conditions for lowering such a symmetry of a BdG Hamiltonian - the result Eq.\eqref{eq:hdelta} is general.

Our work paves the way for further research into spin transport in LaFeO$_3$. A particularly important question that remains is to identify under what conditions dimerization would be achieved in LaFeO$_3$, so that the SNE would manifest via the particular mechanism we discussed above. Another question is whether interesting phenomena might arise in the context of other forms of Eq.\eqref{eq:hdelta}, and in particular whether any such Hamiltonian might arise by accounting for magnon-phonon coupling in bulk LaFeO$_3$ and related materials. 

Of course, when discussing interfacial spin transport the dynamics at the interface can play an important role in addition to the bulk properties we have focused on in this work. Further research might characterize, for example, the magnon dynamics at the interface, which would provide a more complete story regarding the observed magneto-thermovoltage in LaFeO$_3$/Pt.

Having derived a mechanism by which a finite SNE may arise in systems where it is disallowed by symmetry, our work also opens the possibility of engineering transport effects in magnetic systems. For example, this might be accomplished by inducing dimerization as we describe above.

\section{Acknowledgements}
We acknowledge support from
NSF DMR-2114825, DOE Award DE-SC0022168  and the Alexander von Humboldt Foundation.  W.R. acknowledges support from the US Department of Defense through the NDSEG Fellowship.

\bibliographystyle{unsrt}
\bibliography{literature}

\onecolumngrid
\appendix

\section{Bogoliubov-de Gennes Fourier transform and paraunitary transformations} \label{App:BdG}

Expanding around the classical spin configuration above, our Hamiltonian takes the form 

\begin{equation} \label{eq:bosonham}
    H \approx \sum_{\vec{r},\vec{\delta}} \psi^{\dagger}(\vec{r}) H_{\vec{\delta}} \psi(\vec{r}+\vec{\delta}),
\end{equation}
where $\vec{r}$ labels the lattice site and $\vec{\delta}$ are the relevant nearest and next-nearest neighbor separation vectors. Here, $\psi(\vec{r})$ is a Nambu spinor, given by
\begin{equation}
    \psi(\vec{r}) = \begin{pmatrix}
        a_1(\vec{r}) \\
        a_2(\vec{r}) \\
        a_3(\vec{r}) \\
        a_4(\vec{r}) \\
        a^{\dagger}_1(\vec{r}) \\
        a^{\dagger}_2(\vec{r}) \\
        a^{\dagger}_3(\vec{r}) \\
        a^{\dagger}_4(\vec{r})
    \end{pmatrix}.
\end{equation}
We have eliminated linear bosonic terms by expanding around the energetic minimum, and we have dropped constants as well as interaction terms of three bosonic operators an higher, consistent with a Taylor expansion in (1/$S$) of the Holstein-Primakov transformation.

If we let $\vec{x}_{i}$ be the separation between sublattices $1$ and $i$, then we can define a Fourier transformation
\begin{equation}
    a_i(\vec{r}) = \frac{1}{\sqrt{N}}\sum_{\vec{k}} e^{i\vec{k}\cdot (\vec{r} + \vec{x}_{i})} a(\vec{k}).
\end{equation}
Performing this Fourier transform results in the form
\begin{equation} \label{eq:BdG}
    H = \sum_k \psi^{\dagger}_k H_k \psi_k,
\end{equation}
where the $k$-space Nambu spinors are given by 
$$\psi_k = \begin{pmatrix}
     a_1(\vec{k}) \\
        a_2(\vec{k}) \\
        a_3(\vec{k}) \\
        a_4(\vec{k}) \\
        a^{\dagger}_1(-\vec{k}) \\
        a^{\dagger}_2(-\vec{k}) \\
        a^{\dagger}_3(-\vec{k}) \\
        a^{\dagger}_4(-\vec{k})
\end{pmatrix}.$$
The minus sign in the creation operators is inherited from the definition of the Fourier transform at the level of the bosonic operators:
\begin{equation} \label{eq:NambuFourier}
    \psi(\vec{r}) = \frac{1}{\sqrt{N}} \sum_k e^{i\vec{k}\cdot \vec{r}}\begin{pmatrix}
        a_1(\vec{k}) \\
       e^{i\vec{k}\cdot \vec{x}_2}a_2(\vec{k}) \\
        e^{i\vec{k}\cdot \vec{x}_3}a_3(\vec{k}) \\
        e^{i\vec{k}\cdot \vec{x}_4}a_4(\vec{k}) \\
        a^{\dagger}_1(-\vec{k}) \\
        e^{i\vec{k}\cdot \vec{x}_2}a^{\dagger}_2(-\vec{k}) \\
        e^{i\vec{k}\cdot \vec{x}_3}a^{\dagger}_3(-\vec{k}) \\
        e^{i\vec{k}\cdot \vec{x}_4}a^{\dagger}_4(-\vec{k})
    \end{pmatrix} = \frac{1}{\sqrt{N}} \sum_k e^{i\vec{k}\cdot \vec{r}} U_k \psi_k,
\end{equation}
where $U_k$ collects the sublattice separation phases, so that
\begin{equation}
    U_k = \begin{pmatrix}
        1 & 0 & 0 & 0 & 0 & 0 & 0 & 0 \\
        0 & e^{i\vec{k}\cdot \vec{x}_2} & 0 & 0 & 0 & 0 & 0 & 0 \\
        0 & 0 & e^{i\vec{k}\cdot \vec{x}_3} & 0 & 0 & 0 & 0 & 0 \\
        0 & 0 & 0 & e^{i\vec{k}\cdot \vec{x}_4} & 0 & 0 & 0 & 0 \\
        0 & 0 & 0 & 0 & 1 & 0 & 0 & 0 \\
        0 & 0 & 0 & 0 & 0 & e^{i\vec{k}\cdot \vec{x}_2} & 0 & 0 \\
        0 & 0 & 0 & 0 & 0 & 0 & e^{i\vec{k}\cdot \vec{x}_3} & 0 \\
        0 & 0 & 0 & 0 & 0 & 0 & 0 & e^{i\vec{k}\cdot \vec{x}_4}
    \end{pmatrix}.
\end{equation}
This formalism also allows us to relate the position- and momentum-space Hamiltonian matrices $H_{\vec{\delta}}$ and $H_k$. By applying Eq.\eqref{eq:NambuFourier} directly to Eq.\eqref{eq:bosonham}, we find that
\begin{equation}
    H_k = \sum_{\delta} U^{\dagger}_k H_{ \vec{\delta}} U_k.
\end{equation}

Thus, solving the Fourier transformed problem amounts to diagonalizing the matrix $H_k$. However, care must be taken to ensure that the transformation $T_k$ such that $ T^{\dagger}_k H_k T_k = \Lambda_k$ (with $\Lambda_k$ diagonal) also preserves the bosonic commutation relations
\begin{equation}
    [\psi_k,\psi_{k}^\dagger]=\begin{pmatrix}
                     I_4 & 0 \\ 0 & -I_4
                \end{pmatrix}\equiv\eta,
\end{equation}
so that in the diagonal problem $H = \sum_k \Lambda_k \gamma^{\dagger}_k \gamma_k$ the operators $\gamma^{\dagger}_k \gamma_k$ are number operators. It turns out that for this to be the case, $T_k$ must be a paraunitary transformation \cite{10.1093/ptep/ptaa151,doi:10.1146/annurev-conmatphys-031620-104715, PhysRevB.87.174427}, which means
 \begin{equation} \label{eq:paraunitary}
        T_k \eta T^{\dagger}_k = \eta.
    \end{equation}
Eq.\eqref{eq:paraunitary} implies that rather than diagonalize $H_k$ via similarity transformation in the usual way, the eigenvalue problem we should solve to obtain the magnon dispersion is:
\begin{equation} \label{eq:eigen}
        \sigma_3 H_k \ket{\psi^n_k} = E^n_k \ket{\psi^n_k}.
    \end{equation}
The ``kets" $\ket{\psi^n_k}$ that solve the eigenproblem are columns of $T_k$, and from paraunitarity they inherit the normalization condition
    \begin{equation} 
        \bra{\psi^m_k} \eta \ket{\psi^n_k} = \eta_{mn}.
    \end{equation}
We call solutions with norm $1$ ``particle bands" and those with norm $-1$ ``hole bands." Equivalently, the bands with positive eigen-energies (norm $1$) are physical particle bands and the bands with opposite sign are the hole partners.

\section{Generality of symmetry-breaking term} \label{App:HD}

Here we derive the most general form of a symmetry-breaking term that will produce a BdG Hamiltonian with terms like Eq.\eqref{eq:ansatz}. We can look for nearest-neighbor couplings, as these will be the strongest; these couple bosons from neighboring sublattices. Writing the operator content explicitly, and first focusing on pairing terms with we denote with subscript $B$, we can therefore ask about terms such as:

\begin{dmath}
    H^B_{\Delta} = \frac{\Delta^B}{2}\sum_k \left[\sin(kl)b_{-k} a_k - \sin(kl)a_{-k}b_k\right] + {\rm h.c.}
\end{dmath}
where we restrict to one dimension for now for simplicity, since the argument will run the same for both $k_z$ and $k_y$. Here $
\Delta^B$ is an arbitrary, complex constant. The minus signs result from insisting that the ansatz be particle-hole symmetric. Transforming the first term to direct-lattice bosons gives
$$\sum_k \sin(kl)b_{-k} a_k = \frac{1}{2i} \sum_{j} \left( a_j b_{j - x_b -l} - a_j b_{j - x_b + l}\right),$$
where $x_b$ is the direct-lattice separation between the $a$ and $b$ sublattices. From this we see an immediate restriction on the possible values of $l$, stemming from the fact that the index of the $b$ bosons must be a lattice vector; $l + x_b$ and $l-x_b$ must therefore also be lattice vectors. Here $x_b$ is half the length of a unit cell along the direction of interest, and furthermore we wish to restrict to nearest-neighbor couplings. Therefore, the only consistent choices are $l = \pm x_b$. The choice of sign is equivalent to a choice of sign of the overall term, so without loss of generality we can consider the choice $l = x_b$. 

Transforming to spin operators via an inverse Holstein-Primakov transformation, we find that the inversion-breaking term must have the form,
% Below add imaginary pieces I've derived. Same for A 
\begin{equation}
    H^{B}_{\Delta} = \frac{\Delta^B_R}{2i} \sum_j \left[ \tilde{S}^{+}_{aj}\tilde{S}^{+}_{bj-1} - \tilde{S}^{+}_{aj}\tilde{S}^{+}_{bj} -\tilde{S}^{-}_{aj}\tilde{S}^{-}_{bj-1} + \tilde{S}^{-}_{aj}\tilde{S}^{-}_{bj}\right] + \frac{\Delta^B_I}{2} \sum_j \left[ \tilde{S}^{+}_{aj}\tilde{S}^{+}_{bj-1} - \tilde{S}^{+}_{aj}\tilde{S}^{+}_{bj} +\tilde{S}^{-}_{aj}\tilde{S}^{-}_{bj-1} - \tilde{S}^{-}_{aj}\tilde{S}^{-}_{bj}\right]
\end{equation}
$$= \Delta^B_R \sum_j \left[ \tilde{S}^{y}_{aj}\tilde{S}^{x}_{bj-1} - \tilde{S}^{y}_{aj}\tilde{S}^{x}_{bj} +\tilde{S}^{x}_{aj}\tilde{S}^{y}_{bj-1} - \tilde{S}^{x}_{aj}\tilde{S}^{y}_{bj}\right] + \Delta^B_I \sum_j \left[ \tilde{S}^{x}_{aj}\tilde{S}^{x}_{bj-1} - \tilde{S}^{x}_{aj}\tilde{S}^{x}_{bj} -\tilde{S}^{y}_{aj}\tilde{S}^{y}_{bj-1} + \tilde{S}^{y}_{aj}\tilde{S}^{y}_{bj}\right],$$
where $\tilde{S}^{\alpha}$ is a spin operator written in the locally rotated frame. Here we have defined the real numbers $\Delta^B_R$ and $\Delta^B_I$ in $\Delta^B = \Delta^B_R + i \Delta^B_I$.

A similar analysis can be done for non-pairing terms, such as
\begin{dmath*}
    H^A_{\Delta} = \frac{\Delta^A}{2}\sum_k \left[\sin(kl)a^{\dagger}_kb_{k} - \sin(kl)b_{-k} a^{\dagger}_{-k}  \right] + {\rm h.c.}
\end{dmath*}
In this case we find the same restriction on $l$, resulting in possible inversion-breaking terms of the form, 
\begin{equation}
    H^{A}_{\Delta} = \Delta^A_R \sum_j \left[ \tilde{S}^{x}_{aj}\tilde{S}^{y}_{bj-1} - \tilde{S}^{y}_{aj}\tilde{S}^{x}_{bj-1} -\tilde{S}^{x}_{aj}\tilde{S}^{y}_{bj} + \tilde{S}^{y}_{aj}\tilde{S}^{x}_{bj}\right] + \Delta^A_I \sum_j \left[ \tilde{S}^{x}_{aj}\tilde{S}^{x}_{bj-1} + \tilde{S}^{y}_{aj}\tilde{S}^{y}_{bj-1} -\tilde{S}^{x}_{aj}\tilde{S}^{x}_{bj} - \tilde{S}^{y}_{aj}\tilde{S}^{y}_{bj}\right].
\end{equation}
Thus, the form of inversion breaking terms is quite restricted to some linear combination $H_{\Delta} =  H^{A}_{\Delta} +  H^{B}_{\Delta}.$
Letting $\Gamma^{xx} = \Delta^A_I + \Delta^B_I$, $\Gamma^{yy} = \Delta^A_I - \Delta^B_I$, $\Gamma^{xy} = \Delta^A_R + \Delta^B_R$, and $\Gamma^{yx} = -\Delta^A_I + \Delta^B_I$, we can write the term $H_{\Delta}$ as
\begin{equation} 
 H_{\Delta} = \sum_j \Gamma^{\alpha \beta} (S^{\alpha}_{aj}S^{\beta}_{bj-1} - S^{\alpha}_{aj}S^{\beta}_{bj}),
\end{equation}
thus proving that Eq.\eqref{eq:hdelta} is the most general term that breaks inversion symmetry along the direction indexed by $j$. Notably, we find that any choice of couplings $\Gamma^{\alpha \beta}$ will produce symmetry breaking; the key component is the minus sign between the terms inside and outside the unit cell. As is shown in the main text, couplings involving $z$ components do not contribute to the lowest-order Holstein-Primakov expansion and are therefore neglected.

\section{Additional plots} \label{App:Berry}
\begin{figure}[htp]
    \centering
    \includegraphics[width=11 cm]{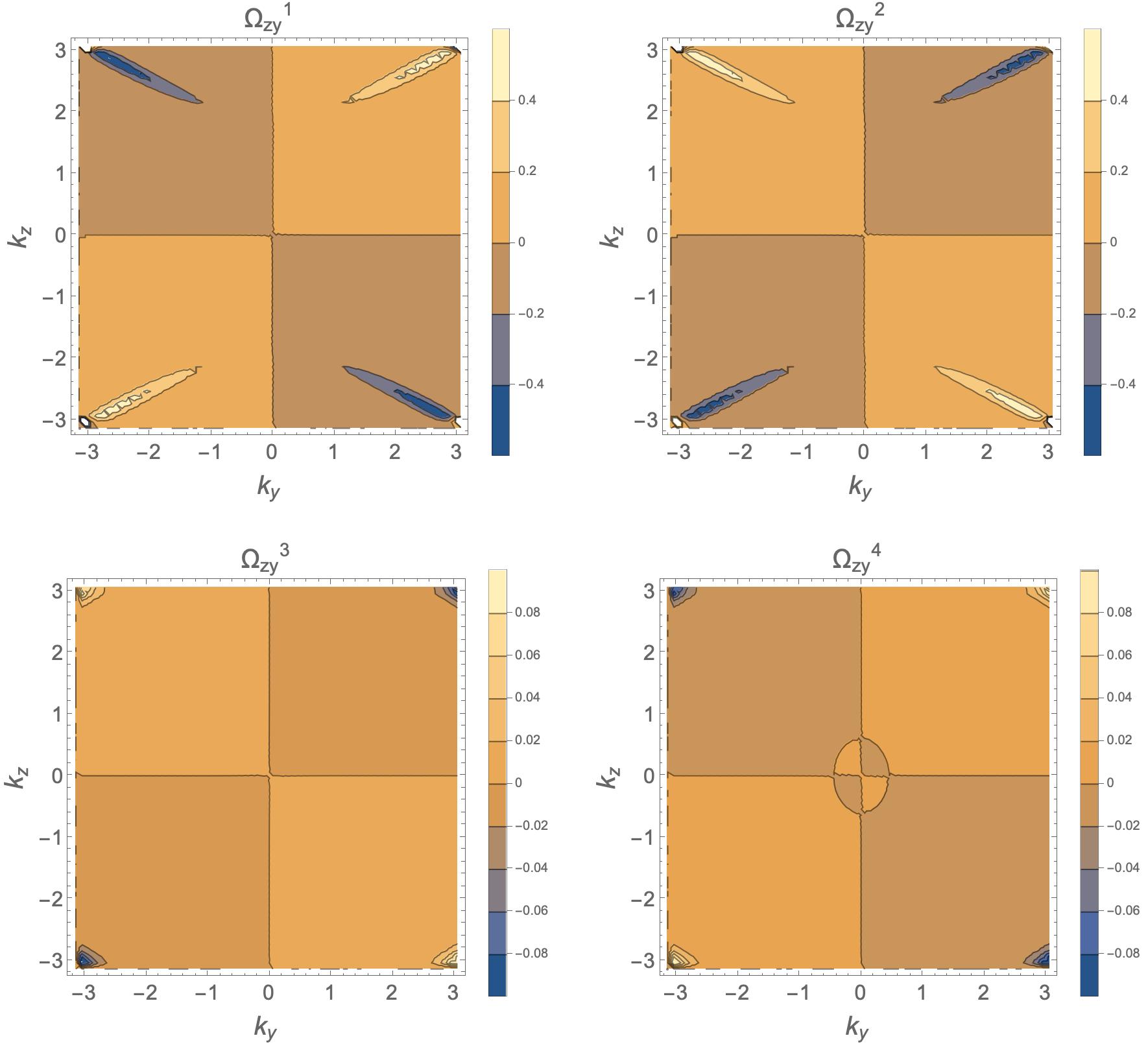}
    
   \caption{\label{Fig:LFOberryfree}Magnon Berry curvature $\Omega^n_{zy}(0,k_y,k_z)$ for LFO with applied field $h = (0,1,1)$, for all bands.} 
    \end{figure}

\begin{figure}[htp]
    \centering
    \includegraphics[width=11cm]{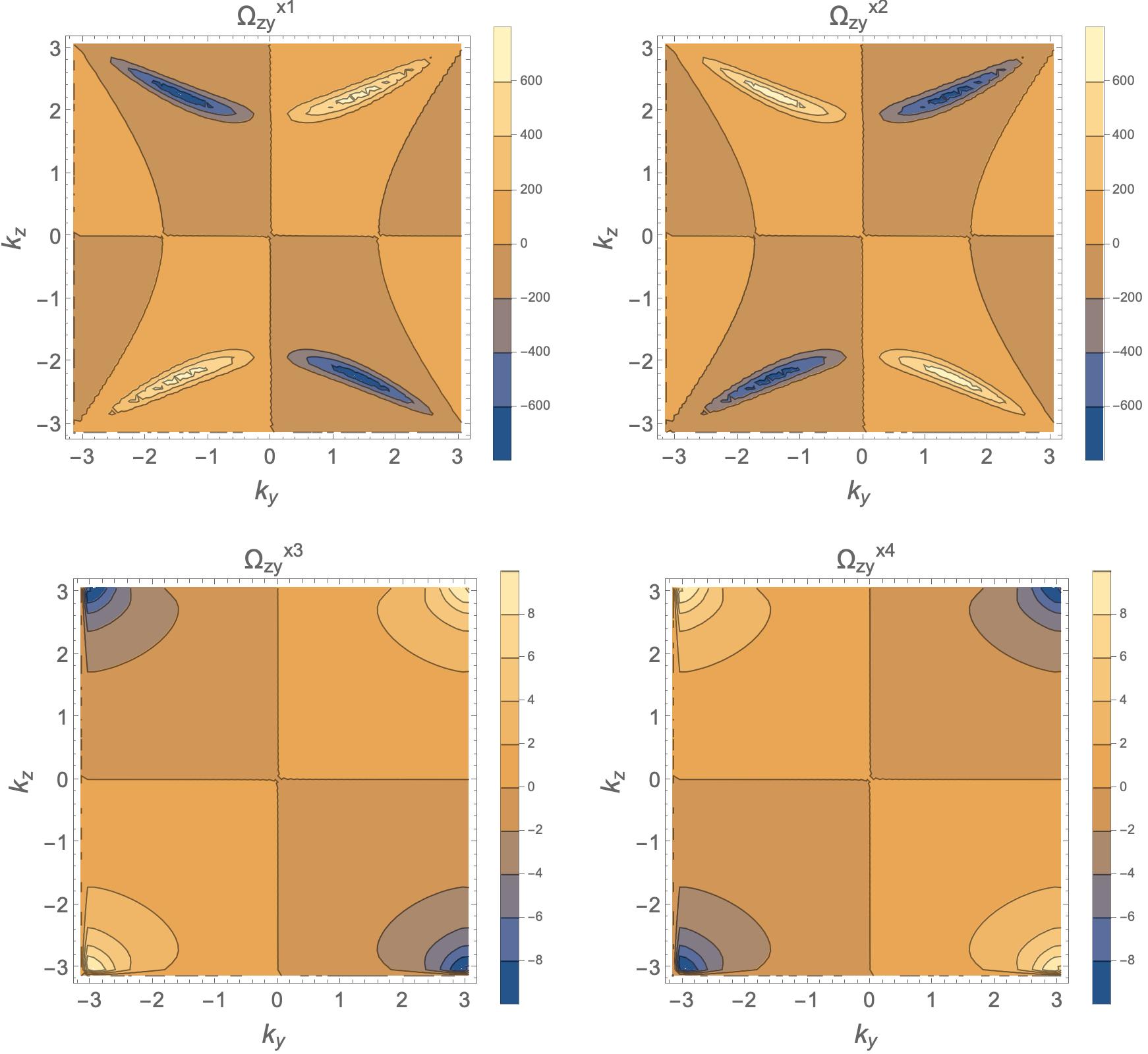}
    
   \caption{\label{Fig:LFOspinberryfree}Magnon spin Berry curvature $\Omega^{nx}_{zy}(0,k_y,k_z)$ for LFO with applied field $h = (0,1,1)$.} 
    \end{figure}

Here we present additional plots to supplement the discussion in the main text. Fig. \ref{Fig:LFOberryfree} displays the band Berry curvature $\Omega^n_{zy}(0,k_y,k_z)$, and  Fig. \ref{Fig:LFOspinberryfree} displays the spin Berry curvature $\Omega^{nx}_{zy}(0,k_y,k_z)$.

\begin{figure}[htp]
    \centering
    \includegraphics[width=13cm]{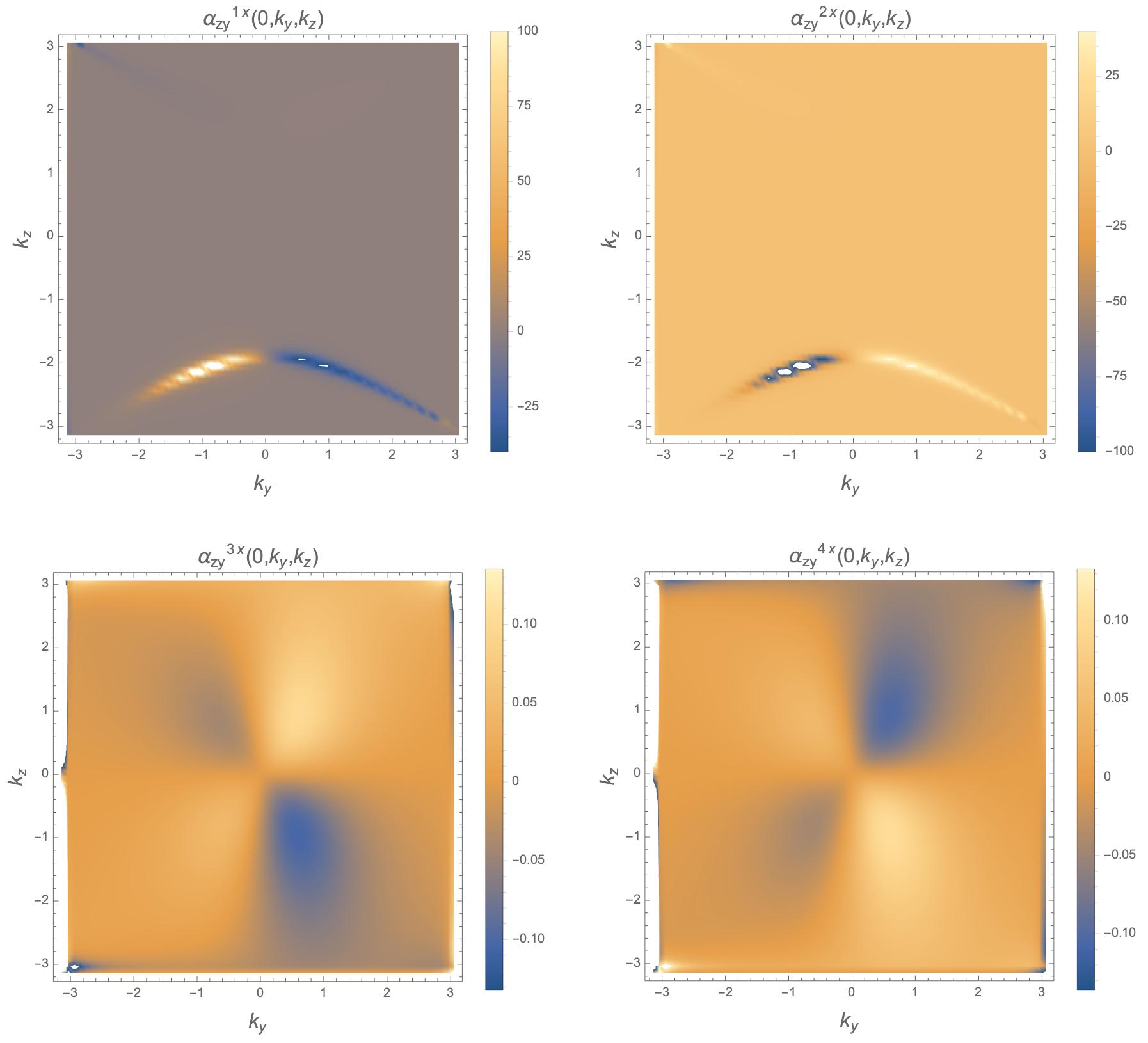}
    
   \caption{\label{Fig:nernstbandwise} The contribution to the SNE from each band, for $k_B T = 10$. This is a low enough temperature that band 2 contributes more heavily than band 1, creating a negative SNE.}
    \end{figure}

    In the main text we considered the net contribution $\alpha^x_{zy}(k_x)$ from all bands. We can also look at the contributions from individual (particle) bands, such that $\alpha^x_{zy} = \sum_n \alpha^{nx}_{zy}$. Here we find that the asymmetrical feature in Fig. \ref{Fig:nernstcoeff} is contributed by the two higher bands, suggesting that to see the effect the temperature must be high enough to allow occupation of these bands. This is depicted in Fig.\ref{Fig:nernstbandwise}.
%Here we see that because $\Omega^{1x}_{zy}(0,k_y,k_z) = -\Omega^{2x}_{zy}(0,k_y,k_z)$, and generally $\Omega^{1x}_{zy}(k_x,k_y,k_z) \approx -\Omega^{2x}_{zy}(k_x,k_y,k_z)$, a net SNE requires differential occupation of these bands.

\end{document}